# NOVEL ASTA USERS FACILITY AT FERMILAB : A TESTBED FOR SUPERCONDUCTING RF TECHNOLOGY AND ERL R&D *

V. Shiltsev# for ASTA team, FNAL, Batavia, IL 60510, USA

*Abstract*

The Advanced Superconducting Test Accelerator (ASTA) currently under commissioning at Fermilab will enable a broad range of beam-based experiments to study fundamental limitations to beam intensity and to develop transformative approaches to particle-beam generation, acceleration and manipulation. ASTA incorporates a superconducting radiofrequency (SRF) linac coupled to a photoinjector and small-circumference storage ring capable of storing electrons or protons. ASTA will establish a unique resource for R&D towards Energy Frontier facilities and a test-bed for SRF accelerators and high- brightness beam applications, including ERLs. The unique features of ASTA include: (1) a high repetition-rate, (2) one of the highest peak and average brightness within the U.S., (3) a GeV-scale beam energy, (4) an extremely stable beam, (5) the availability of SRF and high quality beams together, and (6) a storage ring capable of supporting a broad range of ring-based advanced beam dynamics experiments. These unique features will foster a broad program in advanced accelerator R&D which cannot be carried out elsewhere. Below we describe ASTA and its experimental program, with particular emphasis on the ERL-related accelerator R&D opportunities.

## ACCELERATOR OVERVIEW

The backbone of the ASTA facility is a radio-frequency (RF) photoinjector coupled with 1.3-GHz superconducting accelerating cryomodules (CMs); see Fig. 1-(a) [1]. The electron source consists of a 1-1/2 cell 1.3-GHz cylindrical-symmetric RF gun comprising a $Cs_2Te$ photocathode illuminated by an ultraviolet (UV, 263 nm) laser pulse obtained from frequency quadrupling of an amplified infrared IR pulse The photocathode drive laser produces a train of bunches repeated at 3 MHz within a 1-ms-duration macropulse; see Fig. 1-(b). The 5-MeV electron bunches exiting the RF gun are then accelerated with two SRF TESLA-type cavities (CAV1 and CAV2) to approximately 50 MeV. Downstream of this accelerating section the beamline includes quadrupole and steering dipole magnets, along with a four-bend magnetic compression chicane (BC1) [2]. The beamline also incorporates a round-to-at-beam transformer former (RTFB) capable of manipulating the beam to generate a high transverse-emittance ratio. In the early stages of operation, the bunches will be compressed in BC1. In this scenario the longitudinal phase space is strongly distorted and the achievable peak current limited to less than 3 kA. Eventually, a third-harmonic cavity (CAV39) operating at 3.9 GHz will be added enabling
the generation of bunches with ~10 kA peak currents by linearizing the longitudinal phase space. In addition CAV39 could also be used to shape the current profile of the electron bunch [4]. The photoinjector was extensively simulated and optimized [3]. The photoinjector also includes an off-axis experimental beamline branching at the second dipole of BC1 that will support beam physics experiments and diagnostics R&D.

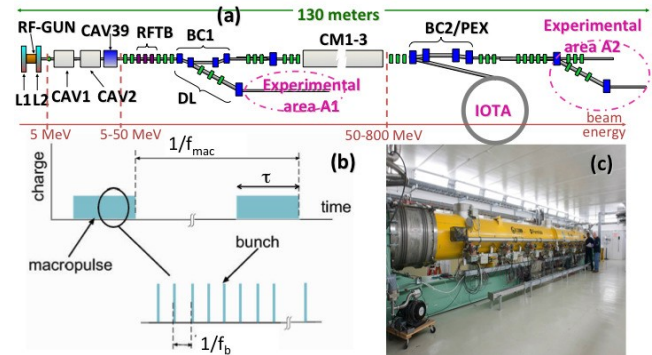

Fig.1: Overview of ASTA (a); "L1" and "L2" stand for solenoids, "CAV1", "CAV2", and "CAV39" correspond to accelerating cavities, "CM1-3" to an ILC cryomodule string, "BC1" and "BC2" to bunch compressors, and "DL" to a dogleg beamline. "EEX" represents a possible reconfiguration of "BC2" to act as a transverse-to-longitudinal phase space exchanger. Electron beam macropulse format (b) and photograph of CM1 module (c).

The 50-MeV beam is injected into the SRF linac, which will eventually consist of three, 12-m long, TESLA/ILC-type CMs. Each CM includes eight 1.3-GHz nine-cell cavities. The first two cryomodules (CM1 and CM2) are a TESLA Type-III+ design, whereas the third (CM3), will be an ILC-Type IV design [5]. Together, these three CM constitute a complete ILC RF Unit. The SRF linac will be capable of generating a beam energy gain of ~750 MeV. The installation of the cryomodules will be staged pending the completion of their construction. The 1st CM has already been tested in the ASTA Facility; see Fig. 1-(c). Downstream of the linac is the test beam line section, which consists of an array of multiple high-energy beam lines that transport the electron beam from the accelerating cryomodules to one of two beam dumps. In addition to testing the accelerator components, the intent of this facility is to also test the support systems required for a future SRF linac. The facility anticipated beam parameters appear in Table I.

___________________
*Work supported by DOE contract DE-AC02-07CH11359 to the Fermi Research Alliance LLC.
# shiltsev@fnal.gov

TABLE I. Beam parameters expected at the ASTA facility; see Fig. 1-(b) for the definitions of the rf macropulse parameters.

| parameter | nominal value | range | units |
|---|---|---|---|
| energy exp. A1 | 50 | [5,50] | MeV |
| energy exp. A2 | ~ 300 (Stage 1) | [50,820] | MeV |
| bunch charge $Q$ | 3.2 | [0.02,20] | nC |
| bunch frequency $f_b$ | 3 | see [a] | MHz |
| macropulse duration $\tau$ | 1 | $\leq 1$ | ms |
| macropulse frequency $f_{mac}$ | 5 | [0.5, 1, 5] | Hz |
| num. bunch per macro. $N_b$ | 3000 | [1,3000][b] | – |
| trans. emittance[b] | $\varepsilon_\perp \simeq 2.11 Q^{0.69}$ | [0.1, 100] | $\mu$m |
| long. emittance[b] | $\varepsilon_\parallel \simeq 30.05 Q^{0.84}$ | [5, 500] | $\mu$m |
| peak current $\hat{I}$ [c] | ~ 3 | $\leq 10$ | kA |

(a) $f_b$ and $N_b$ are quoted for the nominal photocathode laser. Optical pulse stacking methods or field-emission sources could lead to smaller bunch separation within the rf macropulse.
(b) normalized rms values for an uncompressed beam. The scaling laws are obtained from Ref. [3] and correspond to an uncompressed case. Bunch compression results in larger horizontal emittances; see Ref. [2].
(c) the nominal value corresponds to a 3.2-nC compressed bunch without operation of CAV39. Higher values of $I$ are possible with CAV39. For the uncompressed case, we have $I$[A] '=55Q[nC]$^{-0.87}$

During the 1st high energy beam operation and commissioning, only one CM will be installed allowing for the production of bunches with energies up to ~300 MeV. Eventually, the second and third cryomodules will be installed in Stage II. Together, the three cryomodules plus the RF power systems will make up one complete ILC RF unit. During Stage II operation the beam energy will reach approximately 800 MeV. Beyond that stage several options are under consideration, including the installation of a 4th cryomodule downstream of a phase-space-manipulation beamline (either a simple magnetic bunch compressor or a phase space exchanger) [6, 7].

ASTA was designed with the provision for incorporating a small storage ring to enable a ring-based AARD program in advanced beam dynamics of relevance to both Intensity and Energy Frontier accelerators. The Integrable Optics Test Accelerator (IOTA) ring is 39 meters in circumference capable of storing 50 to 150-MeV electrons to explore, e.g, optical stochastic cooling methods [8] and integrable optics [9]. It is planned to expand capabilities for AARD in ASTA by the installation of the 2.5-MeV proton/H- RFQ accelerator which was previously used for High Intensity Neutrino Source (HINS) research at Fermilab [10]. That accelerator starts with a 50-keV, 40-mA proton (or H-ion) source followed by a 2-solenoid low-energy beam transport line. The protons/ions are then accelerated by the pulsed 325-MHz RFQ to 2.5 MeV (with 1 ms pulse duration) prior to injection into IOTA.

Construction of ASTA (a.k.a. NML) began in 2006 as part of the ILC/SRF R&D Program and later American Recovery and Reinvestment Act (ARRA).The facility was motivated by the goal of building, testing and operating a complete ILC RF unit (3 cryomodules). To date, an investment of about $80M has been made, representing ~80% completion of the facility. First beam from ASTA photoinjector has been obtained on June 20, 2013. Pending on available resources, we plan: in FY2013 - to start commissioning of the 50 MeV photoinjector and to install the first user experiments and start the 1.3 GHz SRF cryomodule commissioning; in FY2014 – to carry out the first experiments at 50 MeV, to finish RF (no beam) commissioning of the SRF cryomodule, to install 300 MeV beamline to the beam dump and continue construction of the IOTA ring; in FY2015 – to perform more experiments with 50 MeV and 300 MeV beams and to finish IOTA construction and installation.

Detail description of the facility and proposed experimental program can be found at http://asta.fnal.gov

It was recognized early in the planning process that an e- beam meeting the ILC performance parameters was itself a power resource of interest to the wider Advanced Accelerator R&D community

## USER OPPORTUNITIES

ASTA is intended to be operated 9 months a year as a scientific user facility for advanced accelerator research and development. All the characteristics of a national user facility will be in evidence in the operation of ASTA and its user program. The facility is open to all interested potential users and the facility resources will be determined by merit review of the proposed work. The user program will be proposal-driven and peer-reviewed in order to ensure that the facility focuses on the highest quality research. Proposals will be evaluated by an external Program Committee (the ASTA Program Advisory Committee), consisting of internationally recognized scientists. Proposal evaluation will be carried out according to established merit review guidelines. The 1st PAC and ASTA User's meetings had taken place at Fermilab in July 2013. We expect the first batch of some 30 proposals submitted as part of our proposal to DOE [11] will be reviewed in the Fall of 2013.

Three experimental areas [A1, A2, and IOTA in Fig. 1-(a)] will be available to users for installation of experiments. Area A1, situated in an off-axis beamline within the photoinjector, will provide electron bunches, possibly compressed, with energies up to 50 MeV. The current layout of the off-axis beamline includes a chicane-like transverse-to-longitudinal phase space exchanger, and provision for the installation of a short undulator for beam-laser interaction (e.g., to enable microbunching studies). The high-energy experimental area A2 consists of three parallel beamlines. Two of the beamlines are downstream of doglegs while one is in line with the ASTA linac. Experiments in the three user beamlines and IOTA could be ran simultaneously (switching the beam from one beamline to the other would only require minor optical-lattice adjustment). Finally, the eventual availability of an H- source would allow IOTA to be operated independently of the ASTA electron-beam users.

## ANTICIPATED RESEARCH THRUSTS

*Accelerator R&D for Particle Physics at the Intensity and Energy Frontiers.* The combination of a state-of-the-art superconducting linear accelerator and a flexible storage ring enables a broad research program directed at the particle physics accelerators of the future. The proposed research program includes (1) the test of non-linear, integrable, accelerator lattices (using the IOTA ring) which have the potential to shift the paradigm of future circular accelerator design [12]; (2) the exploration of space-charge compensation schemes in high-intensity circular accelerators, (3) the test of optical stochastic cooling, (4) the investigation of advanced phase space manipulations for beam shaping and emittance repartitioning [13], (5) the exploration of at-beam-driven dielectric-wake_eld acceleration in slab structures [14], (6) the investigation of acceleration and
cooling of carbon-based crystal structures for muon accelerators, (7) measurement of the electron wave function size in a storage ring, (8) high-power target studies for the LBNE experiment, (9) the generation of tagged photon beam for detector R&D, and (10) applications of X-rays produced via inverse Compton scattering to nuclear astrophysics.

*Accelerator R&D for Future SRF Accelerators.* High gradient, high power SRF systems are critical for many accelerator facilities under planning for the needs of high-energy physics, basic energy sciences and other applications. ASTA offers a unique opportunity to explore most critical issues related to the SRF technology and beam dynamics in SRF cryomodules, such as (1) the demonstration of high-power high-gradient operation of SRF CMs with intense beams, (2) the demonstration of technology and beam parameters for the Project X pulsed linac [15], (3) beam-based measurements of long-range wakefield in SRF CMs, (4) ultra-stable operation of SRF linacs using beam-based feedback systems.

*Accelerator R&D for Novel Radiation Sources.* High energy, high-peak and high-average brightness electron beams are crucial to the generation of high-brillance high-ux light sources with photo energies ranging from keVs to MeVs. The high average power and brightness of the ASTA electron beam has unmatched potential for development of several novel radiation-source concepts. Current proposals include (1) the production of high-spectral-brightness X-rays via channeling [16], (2) the generation and application of X-ray using inverse Compton scattering, (3) the generation of narrow-band X-rays [17], and feasibility studies for (4) an XUV free-electron-laser oscillator, (5) the production of attosecond vacuum UV pulses using space-charge-driven amplification of shot-noise density fluctuations, and (6) the investigation of laser-induced microbunching with high micropulse-repetition rate electron beams.

*Accelerator R&D for Stewardship and Applications.* With its high energy, high brightness, high repetition rate, and the capability of emittance manipulations built-in to the facility design, ASTA is an ideal platform for exploring novel accelerator techniques of interest for very broad scientific community beyond high energy physics. Examples of expressions of interest for such explorations include (1) the demonstration of techniques to generate and manipulate ultra-low emittance beams for future hard X-ray free-electron lasers [18], (2) the test of a beam-beam kicker [19] for the Medium-energy Electron-Ion Collider (MEIC) [20] (see also below), and (3) the development of advanced beam diagnostics.

## ERL R&D OPPORTUNITIES AT ASTA

ASTA offers many opportunities for accelerator R&D towards modern ERLs. Among attractive features of the facility are its use of SC RF – the technique of choice for many next generation ERLs (see many contributions to these Proceedings) – and variety of options for beam diagnostics testing. Due to their high average power in CW beams, ERLs need non-intercepting diagnostics of all kinds. Beams of electrons in ASTA with energies of 300-800 MeV and significant charge in long macro-pulses could be a good start for a number of demonstrations, including:
1. CDR for bunch length and beam based feedback at 50 MeV and higher gamma;
2. ODR near field for beam size at 500 MeV and above;
3. OSR from dipoles for beam size and bunch length;
4. EOS based diagnostic;
5. undulator radiation diagnostics (an appropriate undulator is available in ASTA) should be very relevant to ERL) for bunch length, phase, energy, energy spread, beam position, beam size at micropulse level and sub-macropulse;
6. sub-ps source for VUV detector tests;
7. beam-based feedback;
8. beam arrival monitor development;
9. microbunching instability in compressed beams.

One specific experimental proposal for experimental studies at ASTA suggests a proof-of-principle test of the superfast beam-beam kicker, needed for the MEIC collider, currently under development at the Jefferson Lab [20]. The kicker utilizes a low- or medium energy sheet beam of high current density electrons for providing transverse deflection of higher energy electron beam in a ERL system of electron cooling – see Fig.2. This method was initially proposed for two round Gaussian beams [19], but can be significantly expanded to flat beams by using round-to-flat beam transformation beamline available at ASTA.

Experimental studies at ASTA are envisioned in several stages: first, proof-of-principle demonstration of sub-ns EM deflections by external short (round) electron bunches; then, exploration of the efficient and stable deflection with flat bunches; and finally, attainment of the required repetition rate of several MHz, that will probably require energy recuperation of some kind.

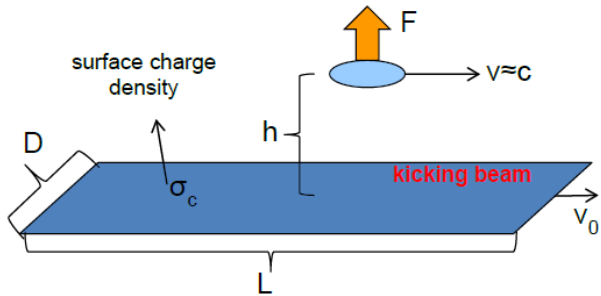

Fig.2: Schematic drawing of a fast beam-beam kicker.

## SUMMARY

Advanced Superconducting Test Accelerator (ASTA) facility is being built and commissioned at Fermilab. It is intended to become a leading US DOE facility for accelerator R&D towards Energy Frontier and Intensity Frontier Accelerators, SC RF developments, novel radiation sources, accelerator stewardship and applications available for domestic and international users. ASTA will be a unique accelerator R&D user's facility because of:
- broad range in beam energies (50-800 MeV)
- high-repetition rate and the highest power beams available
- all the advantages of a modern, SRF-based accelerator
- high beam quality, beam stability, beam brightness
- arbitrary emittance partition with repartition of phase space
- flexible storage ring for novel studies with electrons and protons

ASTA will provide 3 experimental areas for multiple experiments:
- with 50 MeV electrons (Experimental Area 1);
- with 300-800 MeV electrons and SRF (EA2);
- with 50-150 MeV/c electron and protons in IOTA ring (EA3).

The facility offers a number of R&D opportunities for various accelerators, including Energy Recovery Linacs (ERLs), and the ASTA team welcomes new proposals.